\documentclass[journal=nalefd, manuscript=letter]{achemso}

\usepackage[version=3]{mhchem} 
\usepackage{gensymb}

\author{Christopher C. Price}
\email{chris@atomicdatasciences.com}
\affiliation[Atomic Data Sciences]
{Atomic Data Sciences, Boston, Massachusetts 02108, United States}

\author{Yansong Li}
\author{Guanyu Zhou}
\author{Rehan Younas}
\affiliation[University of Notre Dame]
{Department of Electrical Engineering, University of Notre Dame, Notre Dame, Indiana 46556, United States}

\author{Spencer S. Zeng}
\author{Tim H. Scanlon}
\author{Jason M. Munro}
\email{jason@atomicdatasciences.com}
\affiliation[Atomic Data Sciences]
{Atomic Data Sciences, Boston, Massachusetts 02108, United States}

\author{Christopher L. Hinkle}
\affiliation[University of Notre Dame]
{Department of Electrical Engineering, University of Notre Dame, Notre Dame, Indiana 46556, United States}
\email{chinkle@nd.edu}

\title[An \textsf{achemso} demo]
  {Predicting and Accelerating Nanomaterials Synthesis Using Machine Learning Featurization}

\begin{document}


\begin{abstract}
     Materials synthesis optimization is constrained by serial feedback processes that rely on manual tools and intuition across multiple siloed modes of characterization. We automate and generalize feature extraction of reflection high-energy electron diffraction (RHEED) data with machine learning to establish quantitatively predictive relationships in small sets (\~10) of expert-labeled data, saving significant time on subsequently grown samples. These predictive relationships are evaluated in a representative material system (\ce{W_{1-x}V_xSe2} on c-plane sapphire (0001)) with two aims: 1) predicting grain alignment of the deposited film using pre-growth substrate data, and 2) estimating vanadium dopant concentration using in-situ RHEED as a proxy for ex-situ methods (e.g. x-ray photoelectron spectroscopy). Both tasks are accomplished using the same materials-agnostic features, avoiding specific system retraining and leading to a potential 80\% time saving over a 100-sample synthesis campaign. These predictions provide guidance to avoid doomed trials, reduce follow-on characterization, and improve control resolution for materials synthesis. 
\end{abstract}
\section{Introduction}
    
    Differentiated and substantial performance requirements for emerging electronics applications and the deceleration of Moore's Law in silicon is driving demand for advanced materials discovery, optimization, and scale-up \cite{kim_future_2024}. Engineering and development of materials platforms is difficult and time consuming; lab-to-production timelines currently take 10 years or longer, and time-to-market is the primary barrier to commercialization \cite{maine_accelerating_2016}. Significant progress has been made leveraging ab-initio physical simulations (DFT) \cite{hegde_quantifying_2023, yang_big_2022} and subsequent machine learning (interatomic potentials \cite{deng_chgnet_2023, chen_universal_2022, batatia_mace_2022}; generative models \cite{merchant_scaling_2023, zeni_mattergen_2023})  to efficiently identify and screen stable and synthesizable materials candidates in the first stage of advanced materials development. However, the theoretical assumptions of DFT, including the absence of constraints relevant to synthesis in the real world, results in a large time and effort barrier to the realization of materials after promising targets are identified\cite{lee_machine_2022}. While computational capabilities have taken off, working through synthesis recipe design, process optimization, and iteratively improving materials quality relies on a relatively slow manual and intuition-guided experimental approach. To address this bottleneck, recent efforts in both software and hardware have made advancements towards fully autonomous synthesis and optimization within the lab \cite{choudhary_recent_2022, delgado-licona_research_2023, szymanski_toward_2021, xie_toward_2023}. Advanced tools in machine learning and artificial intelligence have proven incredibly useful at targeting both interpretation of experimental data and the subsequent decision making required as part of feedback loop-based solutions, including Bayesian optimization approaches for accelerating search in chemical spaces \cite{szymanski_autonomous_2023, lunt_modular_2024, biswas_dynamic_2024, shields_bayesian_2021, lazin_high-dimensional_2023}. Early versions of these autonomous systems have emphasized the importance and challenges of effective and rapid materials characterization, especially when available datasets are small and the target properties require multiple tools to assess.

    Synthesis optimization is difficult because each trial is time-consuming to conduct and evaluate, especially when nanoscale properties need to be interrogated. While ultra-high vacuum techniques like molecular beam epitaxy (MBE) have highly controlled synthesis environments, the preparation, processing, and subsequent characterization of a single sample takes multiple days \cite{ding_silicon_2016}. Synthesis recipes are highly sensitive, varying across equipment installations and requiring re-calibration after tool maintenance, which can extend over weeks. Due to the expense in time and resources consumed per run, it is critical to maximize the information gained and chance of success for each trial in both manual and autonomously driven settings. In-situ characterization captures large volumes of abstract data with high granularity, yet this data cannot be analyzed with conventional methods in time to impact a trial in progress. An example is reflection high energy electron diffraction (RHEED), frequently used to qualitatively monitor MBE growth \cite{hasegawa_reflection_2012} by providing information on the surface structure of a sample. RHEED images contain a fingerprint of the material surface at a point in time that can take 15 minutes to manually extract for a single image, while processes can change in seconds and data is generated 10 to 100 times per second. Recent work has shown that machine learning can process RHEED data
    \cite{liang_application_2022, vasudevan_big-data_2014, kwoen_classification_2020, gliebe_distinct_2021, yang_https2dmatchemdxorg_2024, provence_machine_2020, kim_machine-learning-assisted_2023}, but these early demonstrations required manual tuning of hyperparameters, fitting to specific materials systems or camera settings, or delivery of results after the run is completed. While providing significant post-run insights, these attributes hinder the general predictive capacity to modify or reduce the number of trials in synthesis optimization, since they require significant system-specific data to be acquired up front.

    In this work, we develop and demonstrate fully automated and general pipelines using both supervised and unsupervised machine learning models to rapidly extract physically-motivated and holistic quantitative fingerprints from RHEED data. We show that these fingerprints can speed up the synthesis feedback loop by constructing predictive models from small datasets ($\sim$10 samples) of labeled trials to provide relevant feedback from ex-situ analysis using only in-situ inputs. These predictive models are demonstrated in two stages of the synthesis process for the target system, two-dimensional (2D) \ce{V}-doped \ce{WSe2} on \ce{Al2O3}(0001) (sapphire): 1) evaluating the probability of a substrate to produce grain-aligned film growth, and 2) estimating the composition of a dopant in the film before ex-situ x-ray photoelectron spectroscopy (XPS) is conducted. For both objectives, the success of the predictive models can save significant time and cost by avoiding doomed trials and reducing the number of steps required to assess the sample. By producing these empirically derived predictions for near-real-time feedback, we show that the synthesis optimization loop can be accelerated, leading to higher throughput of material samples with the target characteristics.

\section{Results and Discussion}

\begin{figure}
  \includegraphics[width=1.0\textwidth]{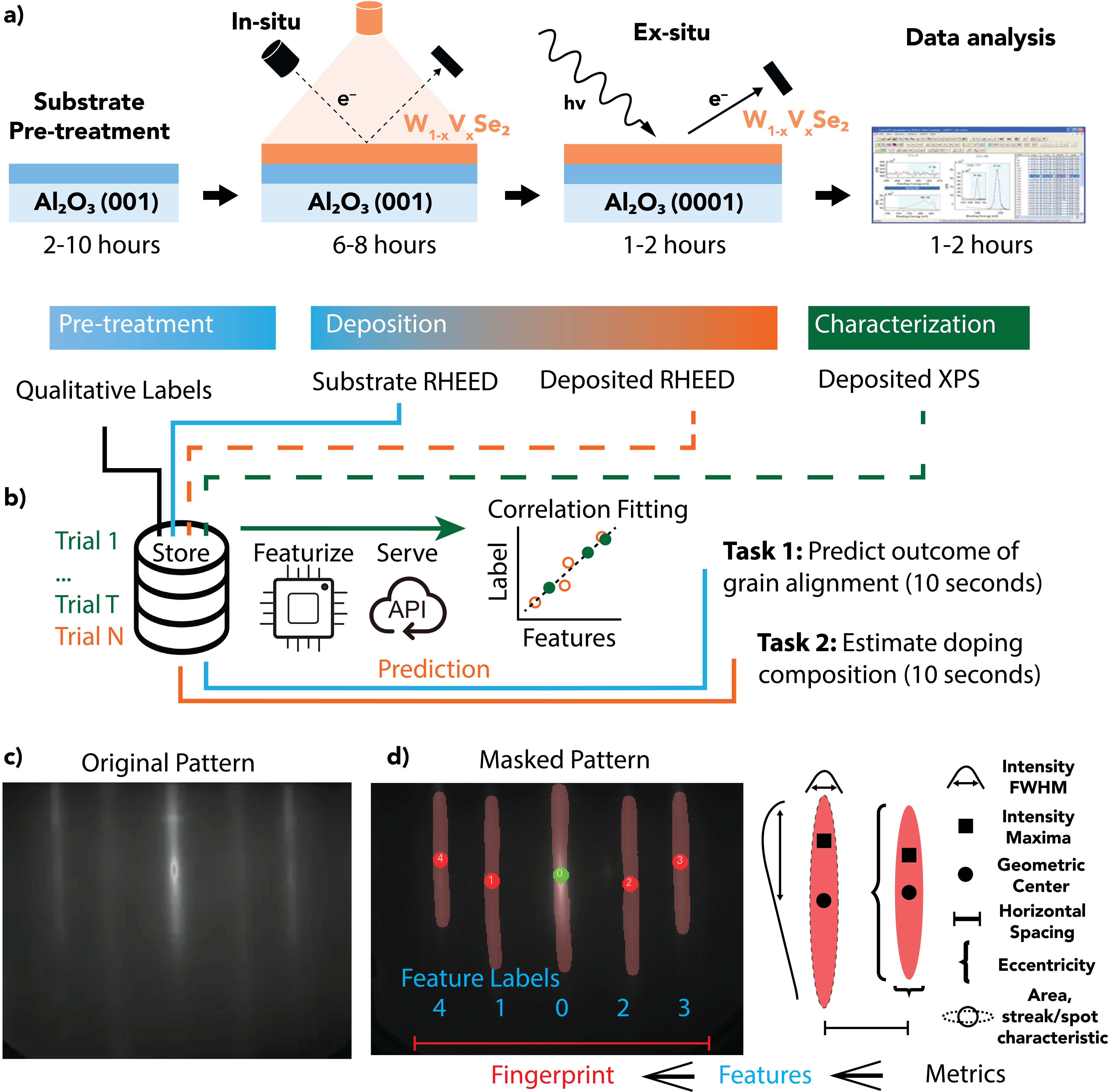}
  \caption{(a) Summary of experimental flows for sample preparation, film growth, and characterization. At the beginning and end of MBE deposition, in-situ RHEED is collected and automatically fingerprinted. After synthesis, the sample is transferred for XPS characterization. (b) Summary of data analysis flows for synthesis and characterization data. Labeled trials are iteratively updated in the database, and correlation fitting is performed for the two tasks against the input labels. Next-trial predictions are generated within 10 seconds. (c) An image of a RHEED pattern of the as-grown film, and (d) the color mask representing featurized regions. Comprehensive metrics are extracted for each diffraction feature to form a complete fingerprint unbiased by user priors. Fingerprints are input to the empirical correlation models; see supporting information (SI) section 1.}
  \label{figure1}
\end{figure}

    Precise doping control in 2D materials is a difficult but necessary milestone to achieve for the next generation of power and space efficient semiconductors \cite{younas_perspective_2023}. Vanadium doping in \ce{WSe2} gives p-type doping with spin polarization, making this system a candidate as a high-mobility dilute magnetic semiconductor \cite{zhang_monolayer_2020,yun_ferromagnetic_2020}. It is crucial to control the dopant concentration during co-deposition of V and W to minimize domain formation and phase separation, key challenges in this materials system. Illustrating the limitations of theoretical prediction, reliably synthesizing theoretically stable, uniformly distributed phases requires high-fidelity control of the kinetics connecting the input procedure to the resulting material composition and microstructure. End-to-end, each trial to map out these relationships can consume over 24 combined hours of tool and active operator time, even excluding the time required for sample loading and equipment standby (Fig.  \ref{figure1}a). Here, we develop a general framework that can be used to avoid doomed trials and map ex-situ measurements to in-situ characterization in molecular beam epitaxy to save 80\% of the time over a 100 trial synthesis campaign. 
\subsection{Automated Data Workflows in the Synthesis Context}

    The data workflow for automated generalized characterization analysis is given in Fig. \ref{figure1}b. The input RHEED images are passed through a featurization pipeline which extracts, normalizes, and labels diffraction features categorized in Fig. \ref{figure1}d. Images are first cropped to remove artifacts from the detector and an image segmentation pipeline composed of two models -- a U-Net architecture for RHEED proposed by Liang \textit{et al.} \cite{liang_application_2022} followed by a transformer-based segmentation model \cite{kirillov_segment_2023} tuned for performance on low-contrast medical grayscale images \cite{ma_segment_2024}. Output masks from this segmentation pipeline are labeled to identify contiguous diffraction regions and comprehensive metrics are computed for each diffraction feature (Fig. \ref{figure1}d). A coordinate system using the specular spot, identified by relative position to the other features, as the origin is adopted to enable comparison of diffraction features across different patterns. The original RHEED pattern in Fig. \ref{figure1}c shows typically diffuse scattering features that need to be consistently separated from the background, highlighting the challenges of manual analysis and the need for task-specific models.  In the featurization scheme, no hyperparameters are input or adjusted across different patterns or materials systems to maximize generalizability of the workflow and enable real-time result generation without operator intervention.
   
    These automatically generated diffraction fingerprints are correlated with qualitative labels on grain orientation and the quantitative results of manual XPS analysis in an effort to 1) predict whether a growth is likely to lead to aligned or randomly oriented grains (textured growth) based on the RHEED image of the substrate wafer before deposition starts, and 2) estimate the Vanadium doping composition in a deposited film using a RHEED snapshot as input.  These tasks were designed to solve real challenges encountered over a three-year period of aiming to synthesize high quality samples.  The automated process takes 10 seconds to produce featurized RHEED datasets per frame and 10 seconds to generate predictions from the correlative models derived from the task specific training samples, significantly shortening the feedback time relative to traditional approaches; details in the \textbf{Methods} section.

\subsection{Independent Classification of Film Crystallinity from Film and Substrate RHEED}

\begin{figure}
  \includegraphics[width=1.0\textwidth]{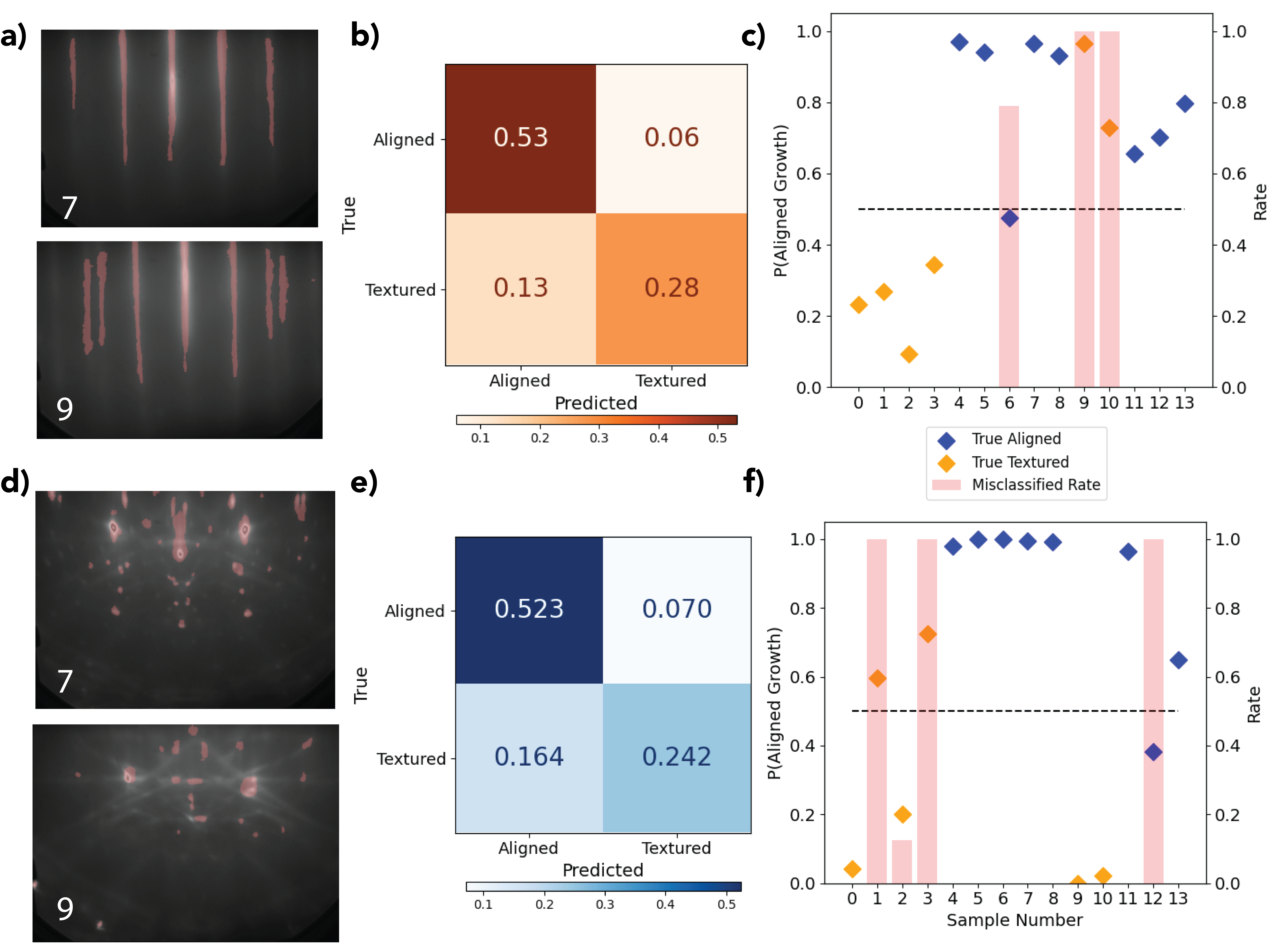}
 \caption{(a) Segmented RHEED patterns for examples of aligned (top) and textured (bottom) \ce{WSe2} film growth. Labels in the bottom left correspond to sample number. (b) Confusion matrix and classification accuracy for a logistic regression model fit with bootstrap aggregation to a set of 14 samples of featurized \ce{WSe2} patterns. (c) Probability of aligned growth predictions by sample (scatters) and frequency of misclassification (bars) for the \ce{WSe2} RHEED data. (d) Segmented RHEED patterns for examples of sapphire substrates that led to aligned (top) and textured (bottom) film growths. (e) Confusion matrix and classification accuracy for the same model structure in (b), fit to the substrate RHEED instead of the film RHEED against the film labels. (f) Same as (c) for the sapphire substrate pattern classification task.
  }
  \label{figure2}
\end{figure}

    Given the critical role of crystallinity in material synthesis and downstream device performance, identifying whether a deposited film has aligned grains can avoid doomed efforts that lead to low quality samples and wasted time. In the case of 2D chalcogenide growth, epitaxial alignment is particularly susceptible to the surface topography of the sapphire substrate \cite{mortelmans_peculiar_2019}. Surface reconstructions ($1$x$1$ Al-terminated or $(\sqrt{31}$ x $\sqrt{31})R9$ ) supporting aligned growth are achieved by thermal annealing, but significant variance in the results exists due to coupling of the annealing procedure with the individual wafer and furnace conditions.  Fig. \ref{figure2} shows the classification results based on the featurized RHEED datasets and an initial label set from visual inspection which categorizes as-grown films as either textured or aligned. Fig. \ref{figure2}a shows examples of strongly aligned (top) and strongly textured (bottom) \ce{WSe2} films, and Fig. \ref{figure2}b-c gives the baseline classification results for the deposited \ce{WSe2} films; details of classification are given in \textbf{Methods}. We restrict to small training datasets to mimic the typical data availability in the early stages of a synthesis effort and maximize the ability to provide guidance for a subsequent trial. The confusion matrix in Fig. \ref{figure2}b gives the binary grain alignment prediction accuracy of 80\%; further details of the bagging procedure are given in \textbf{Methods}. The classification probabilities for grain-aligned films are plotted in Fig. \ref{figure2}c along with the misclassification frequency for each sample. The probability of classification serves as an uncertainty metric and a quantitative approximation for the degree of overall grain alignment. Some samples are always misclassified when held out of the training set, including the canonically textured film 9; explanation for this is given in SI section 2. Overall, the RHEED features contain enough signal to automatically match the expert-identified trends in the labels with a small set of examples. Automating this task removes operator bias from data analysis, and quantification helps set thresholds for films which meet the quality criteria for subsequent device fabrication. However, additional operator time, tool time, and resources could be saved by avoiding low quality film growths before they occur. 

    Natural variance with substrates and precursors can lead to unexpected growth outcomes, even if the same recipe is programmatically followed; in this \ce{WSe2} system deposition alone consumed up to 8 hours, and average all-in synthesis times can range from 5 to 14 hours for MBE \cite{he_molecular_1999}. Using the same data infrastructure and feature extraction as for the \ce{WSe2} films, we perform an identical fitting procedure using the pre-deposition sapphire RHEED patterns as inputs instead of the as-grown films, with results in Fig. \ref{figure2}d-f. Fig. \ref{figure2}d similarly shows the different surface reconstructions of the sapphire that can lead to aligned (top) or textured (bottom) growth. The logistic regression classifier with bootstrap aggregation achieves an accuracy near 80\%, similar to the results from the \ce{WSe2} RHEED, although the most-misclassified samples differ than those using the film RHEED. For the misclassified samples in Fig. \ref{figure2}c, the quantitative probability score is close to 50\%, indicating greater uncertainty of prediction. We show that the quantitative classification probability for both datasets correlates with qualitative assessment with a detailed view of sample 12 in Fig. S1; this sample is labeled as grain-aligned but shows a lower classification probability for both the substrate and the film. In the sapphire RHEED data for sample 12, several features of the $(\sqrt{31}$ x $\sqrt{31})R9$ reconstruction are missing compared to the aligned-producing substrate 7, and the Kikuchi lines are better matched with the textured-generating substrate of sample 9. In the as-deposited film, the pattern is consistent with an aligned growth, but there are identified small features that are signatures of the textured films. This indicates that the classification probability contains information about the quantified likelihood of grain alignment in the deposited film conditional on the substrate.  This provides a new resource for deciding whether to proceed with a growth on a given substrate; if the likelihood of achieving high-quality growth is deemed low, operators have the option to perform additional pre-growth treatment or switch to a different substrate, rather than proceeding with a likely doomed trial.

\subsection{Mapping Ex-Situ Measured Composition to In-Situ RHEED}

\begin{figure}
  \includegraphics[width=1.0\textwidth]{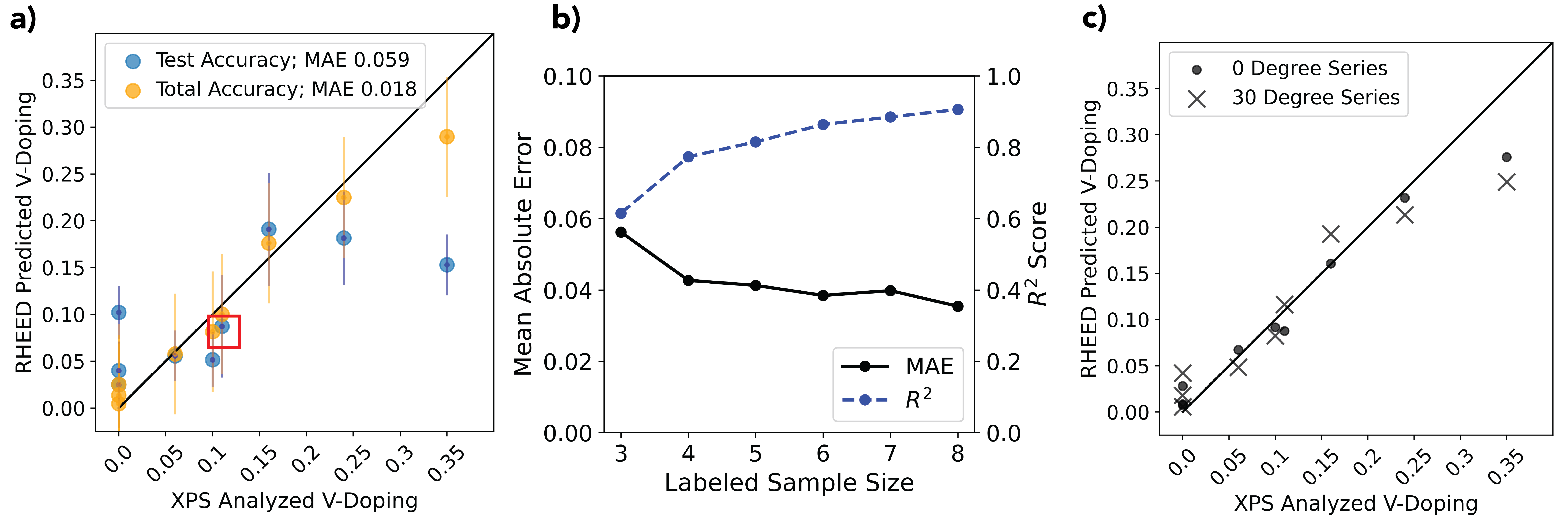}
 \caption{(a) Plot of predicted vs. actual vanadium doping composition \ce{W_{1-x}V_xSe2} assessed by XPS measurement (x-axis) and predicted from RHEED features (y-axis). Orange points show the predictions from a model fit to all 9 data points indicative of overall correlation; blue points show the composition prediction for each data point from a model generated with that point withheld. Predictions are the average of models independently fit to 0$\degree$ and 30$\degree$ data series. The black line is a visual guide to indicate zero absolute error between the XPS-derived composition and the RHEED-predicted composition result. Error bars give the standard deviation of predictions for the individual estimators within the bagging ensemble; MAE is mean absolute error of $x$. (b) Monotonic improvement in prediction accuracy for composition with added training samples, indicating avoidance of overfitting and tunability of desired prediction precision (c) Predictionsdseparately generated for the two independent RHEED series collected on the same samples at two different azimuthal angles separated by 30$\degree$ (dots and x's). Averaging the prediction at each labeled composition gives the orange points in (a). }
  \label{figure3}
\end{figure}
    After improving the yield of films with the target grain microstructure, we turn to optimization and control of film composition by estimating dopant concentration while the sample remains in the growth chamber. Conventionally, dopant concentrations are determined ex-situ by characterization techniques like XPS and energy-dispersive x-ray spectroscopy (EDX) after the entire growth session is complete, requiring time-consuming sample relocation. The scattering factors of different elements create intensity modulations and shifts in the RHEED pattern, but this information is difficult to assess directly since it is tightly convolved with other diffraction mechanisms \cite{peng_electron_1998, kawamura_finding_2022, pawlak_analysis_2021}. The relationships between composition, tool parameters, and growth recipe are nonlinear even for highly controlled synthesis environments and fully mapping design space requires extensive trial and error. If compositional feedback can be quickly generated and delivered from in-situ inputs, better control can be exerted over compositional doping and process refinement.
    
    Fig. \ref{figure3} gives prediction results for V-doping composition based on automatically generated RHEED features; predictions are averaged from the output of two separate models fit on two azimuthal angles collected for each model system (\ce{W_{1-x}V_xSe2}) sample. By labeling a small set of RHEED images with composition from XPS analysis of the \ce{V}-2p peak (Fig. S3), we show that substantial predictive capability can be uncovered from a small initial dataset. Due to the small number of labeled samples, we apply linear regression with bagging to look for correlations between the featurized RHEED and the XPS quantification; details in \textbf{Methods}. Interpolative test accuracy is strongest as shown by the blue test predictions in Fig.\ref{figure3}a, and aside from the composition endpoint ($x=0.35$) of the target range there are no substantially outlying predictions. This highlights the importance of combining data-driven practices with experiment design to maximize the strength of the surrogate models: given a set of labeled examples, the surrogate models provide accurate interpolative estimates for new samples. The red square in Fig. \ref{figure3}a highlights a sample synthesized and characterized by a different operator months after the initial campaign that spanned multiple tool maintenance cycles, demonstrating the persistence of the identified trend to factors that affect the consistency of growth; details in Fig. S2. Fig. \ref{figure3}b shows the prediction accuracy improving monotonically with each additional labeled training point in both mean absolute error (MAE) and coefficient of determination ($R^2$), indicating the correlation model is not overfit. Fig. \ref{figure3}c shows close agreement in predictions generated independently from two high-symmetry azimuthal angles, which acts as a physical sanity check and emphasizes that the V-doping correlated scattering changes are not an artifact of data collection. A full accounting of features input to the composition regression along with their Pearson correlation coefficients for model interpretability is given in Fig. S4. The coefficients show that the first order features contain the most correlated variance in the 0-degree azimuthal data, while the second-order features contain more variance in the 30-degree series. The full-width half-maximum and feature axis lengths have the greatest correlation magnitudes, indicating that the shape of the internal diffraction intensity distribution is most correlated with the compositional change. These data properties are difficult to assess visually even for experts without automated tools. Individual metric analysis serves as an entry point for deeper physics-based analysis, indicating atomistic mechanisms that may be correlated with the target property.

\section{Conclusions}

    We demonstrate that machine learning models tailored for RHEED data can extract high-fidelity feature sets that reveal rich relationships across materials systems, even in the limit of small sets of labeled data. Predictions based on these relationships can help avoid synthesis trials with high failure probability, reduce the amount of ex-situ characterization required, and provide real-time feedback on properties traditionally only measured ex-situ. With research projects needing hundreds of samples and material processing requiring thousands, delivering these predictions with dedicated data infrastructure could save thousands of expert hours in preparation and analysis. Our approach complements intelligent experiment design algorithms for synthesis, such as Bayesian optimization, by accelerating the acquisition function and providing higher-quality inputs for adaptive search.

\section{Methods}

\subsubsection{Sapphire Substrate Treatment}
    2-inch \textit{c}-plane sapphire substrates (0001) (Cryscore) with intentional miscut of C off A by 0.2$\degree$were pre-growth annealed at 1200 $\degree$C for 5 hours in a tube furnace (Thermco Furnace) inside a class 100 cleanroom. During the annealing, the nitrogen gas continuously flowed at 3 L/min through the furnace under atmospheric pressure.

\subsubsection{Vanadium-doped WSe\textsubscript{2} Thin Film Growth}
    The \ce{W_{1-x}V_xSe2} thin films were grown on the pre-treated \textit{c}-plane sapphire substrates (Cryscore) in an MBE system (Scienta Omicron) with a base pressure of \~10\textsuperscript{-10} mbar. Prior to the deposition, the sapphire substrate was degassed at 900 $\degree$C for 1 hour. Then the thin films were grown at 550 °C by co-depositing V  from an effusion cell, Se evaporated from an effusion cell with a cracker zone, and W from an electron-beam evaporator. A relatively high Se flux of 1.0×10\textsuperscript{-6} mbar, measured by beam flux monitor, was used to minimize Se vacancies in the films. After the co-deposition, the samples were still exposed to the same Se flux and annealed at 700 °C for 2 hours. To monitor the growth process, in-situ RHEED (Scienta Omicron) was operated with an acceleration voltage of 13 kV and a current of 1.48 A. 

\subsubsection{X-ray Photoelectron Spectroscopy}
    After growth, the samples were transferred directly to the integrated XPS chamber without breaking the vacuum. XPS measurements were carried out using an Al K-$\alpha$  X-ray source (1486.7 eV) under a background pressure of 2.6×10\textsuperscript{-9} mbar. An electron neutralizer was turned on during the measurement to compensate for charging effects caused by the insulating sapphire substrate. 

\subsubsection{Data Analysis}
    RHEED featurization datasets were generated using AtomCloud RHEED analysis software and accessed via API for training and inference. Correlation models are taken from the scikit-learn python package. Principal component analysis (PCA) is performed on the training dataset for substrate classification and inference only, and bootstrap aggregation with 100 estimators is used with a logistic regression base estimator. Metrics are averaged over 128 independent trials with different held out data. The same fit PCA components are used to transform data for inference. For dopant composition estimation, no PCA preprocessing is used before bootstrap aggregation around the linear regressor. PCA is used with the sapphire featured dataset because of substantially higher number of distinct diffraction features; PCA preprocessing compression gives more stable results in the logistic regression fitting. Evaluation was done to best simulate the accuracy of the next random prediction given a random prior distribution, including class imbalance. Classifiers are never averaged across model runs in evaluation to avoid test leakage. For task 2 (XPS regression), the same fitting procedure is followed using 50 bagging estimators, due to the reduced combinatorics of the input dataset. Jupyter notebooks used for the correlative modeling are available with live data integration at https://github.com/atomic-data-sciences/api-client/tree/main/examples.

\begin{suppinfo}
Explanation of RHEED feature set; example of mixed grain aligned substrate and film RHEED;  forward predictive inference highlighting model utility; RHEED feature label scheme and Pearson correlation coefficients between features and V-composition.
\end{suppinfo}
    

\begin{acknowledgement}
This work was supported in part by SUPREME, one of seven centers in JUMP 2.0, a Semiconductor Research Corporation (SRC) program sponsored by DARPA. This work was also supported in part by the DMREF program of the National Science Foundation (NSF) through the Division of Materials Research (DMR) Awards No. 2324172 and 1921818. 
\end{acknowledgement}

\bibliography{manuscript}

\begin{figure}
    \includegraphics[width=0.5\linewidth]{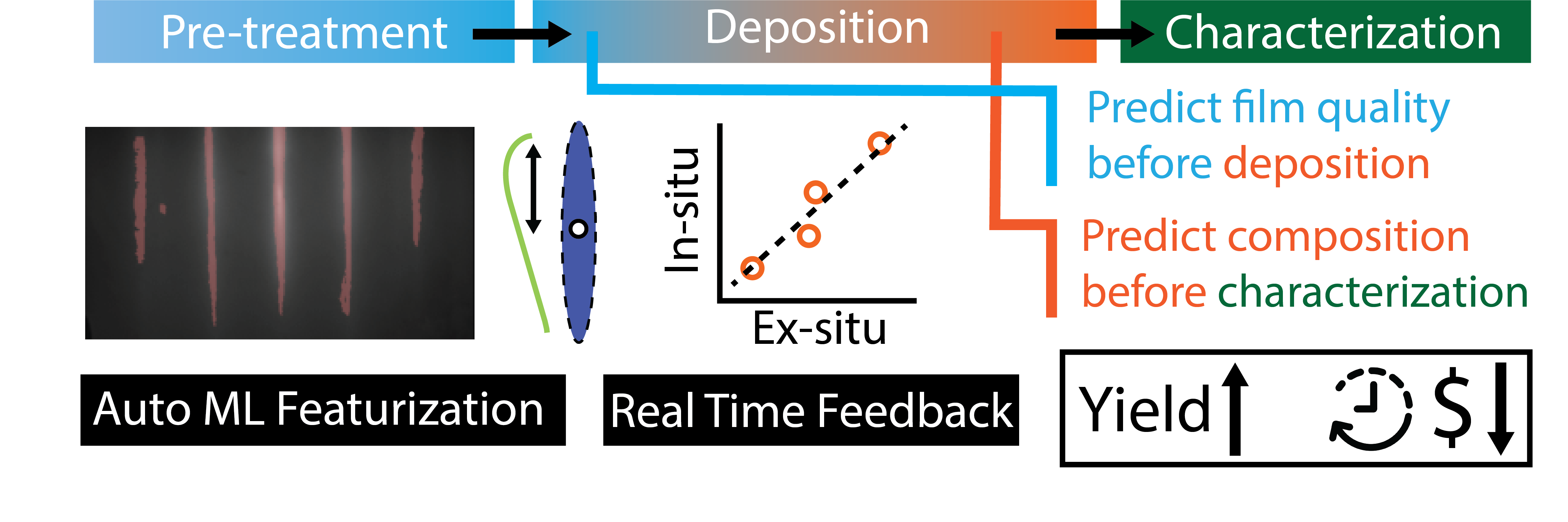}
    \label{toc}
    \caption*{For Table of Contents Only}
\end{figure}

\end{document}


\section{1. Physical Attributes of Feature Regions}

    Feature regions in electron diffraction patterns are quantitatively characterized as comprehensively as possible in an effort to capture as much of the physically relevant diffraction signal in a format compatible with correlative modeling and to minimize operator bias. The specific metrics captured for each feature in a pattern are listed below. Common physical interpretations of each metric are also noted:

\begin{enumerate}

\item Intensity full width half max (FWHM) along the major and minor axis of the feature ellipsoid.
\item Intensity maxima position - 2D coordinates
\begin{enumerate}
    \item (1,2) together indicate surface roughness and domain size in some cases \cite{hafez_review_2022}.
\end{enumerate}
\item Geometric center position - 2D coordinates
\item Total spacing between the geometric centers of neighboring features. 
\begin{enumerate}
    \item (3,4) contain embedded information about surface symmetry as well as in-plane lattice parameter \cite{hasegawa_reflection_2012}.
\end{enumerate}
\item Horizontal spacing between the geometric centers of neighboring features. 
\begin{enumerate}
    \item Proportional to the surface-plane lattice spacing perpendicular to the RHEED incident beam azimuth.
\end{enumerate}
\item Major and minor ellipsoid axis length and elongation (ratio of axes). 
\item Total diffraction area, diffraction area classified as streaks, diffraction area classified as spots, adapted from \cite{liang_application_2022}
\item Streak to spot area ratio. 
\begin{enumerate}
    \item Indicator of the surface morphology, where spots + diffractive arrangement indicate single crystal and streaks + diffractive arrangement indicate morphologically smoothly surfaces with some disorder \cite{mahan_review_1990, hasegawa_reflection_2012}.
\end{enumerate}
\end{enumerate}

\section{2. Classification of Film Crystallinity}

    In the qualitative film classification task, some samples are always misclassified when held out of the training set, including the canonically textured film 9. This is a limitation of the small dataset size; when film 9 is not included in fitting, the signal identified to correlate with textured films changes significantly and the classifier ends up in a different local minima. This highlights the importance of including strong examples of target phenomena in the training set, and that the value this approach lies in interpolating between bounds and disambiguating examples which fall between the range of what is clearly visually identifiable. 
\begin{figure}
  \includegraphics[width=0.5\textwidth]{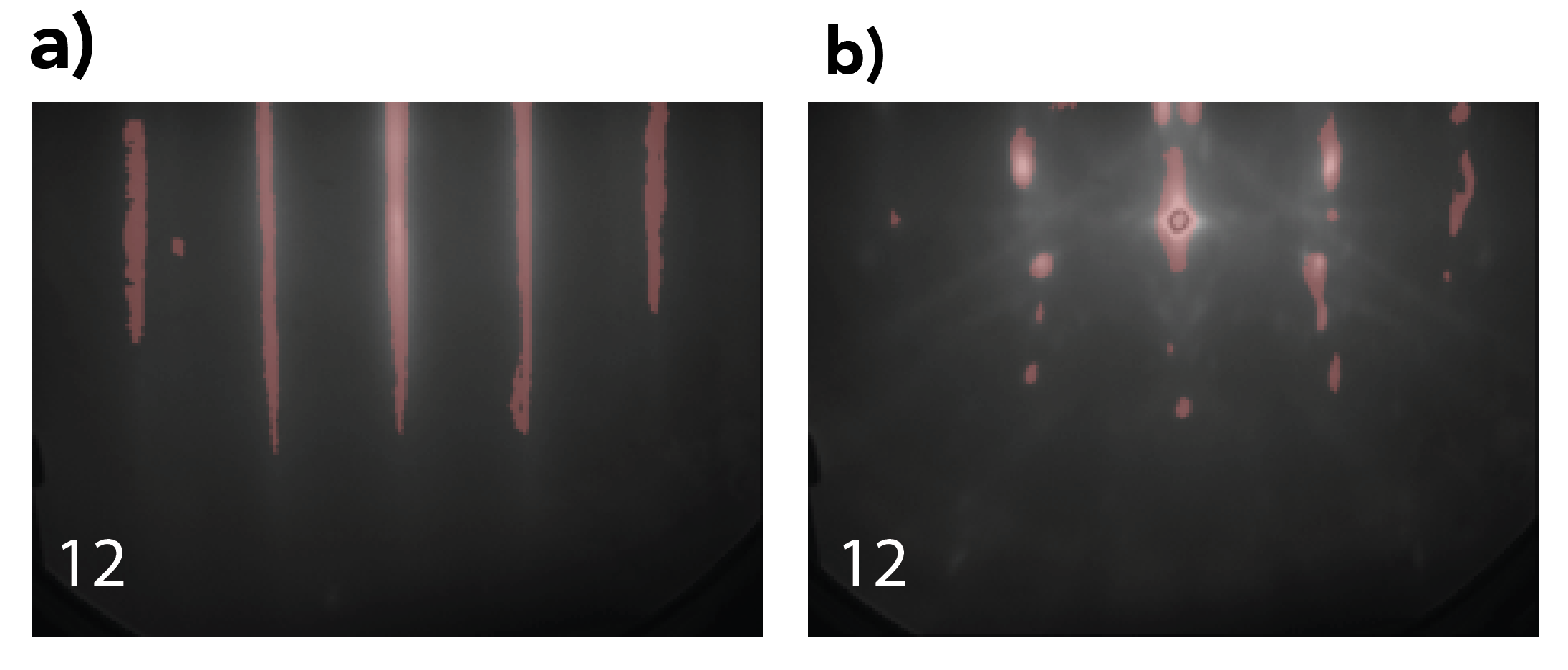}
 \caption{Zoomed-in view of an aligned growth which demonstrated a lower probability of classification. The substrate RHEED (b) shares features of both strongly predicted aligned and textured examples in Fig. 2d and small texturing features are visible in the film RHEED (a), consistent with the lower probability of the aligned prediction.
  }
  \label{figureS1}
\end{figure}

\section{3. Predictive Sample Inference}
    
\begin{figure}[!htb]
  \includegraphics[width=1.0\textwidth]{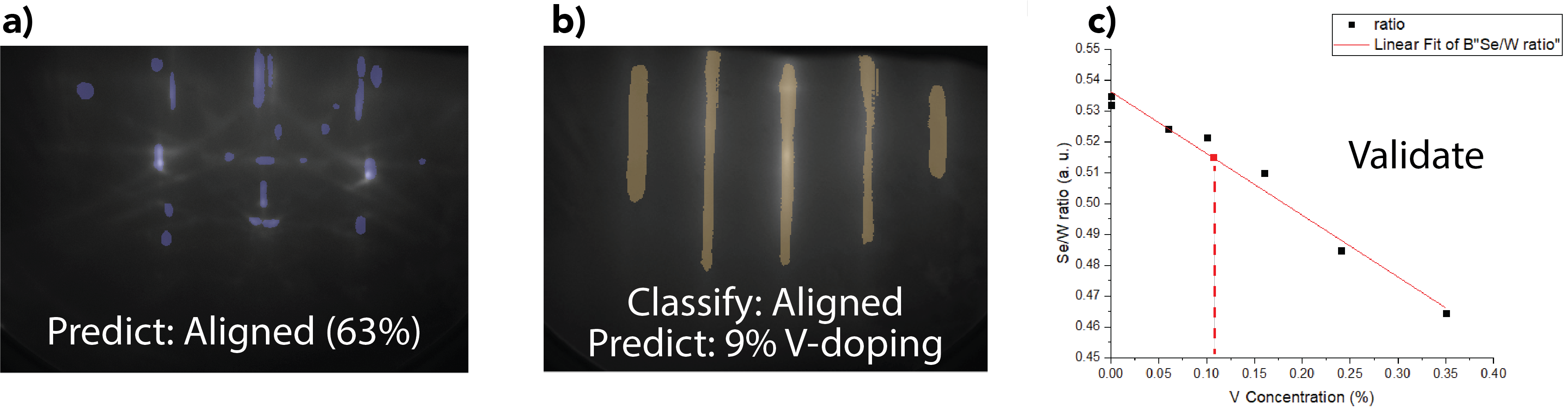}
 \caption{(a) Inference results using the models from Fig. 2 and Fig. 3c on the most recent set of substrate and film data, collected 3.5 years after the earliest sample in the training set.
  }
  \label{figureS2}
\end{figure}

    We tested the forward generalizability of all three inference models on a freshly prepared \ce{W_{1-x}V_xSe2} sample grown 3 months after the latest sample data and 3.5 years after the earliest sample data used to develop the models in this study, without retraining the core featurization scheme or any of the correlative models (Fig. \ref{figureS2}). This removes any implicit bias introduced in the train, test, and hyperparameter optimization, and extends the method application to clean synthesis. From the substrate RHEED data alone, our inference scheme predicted that aligned growth would be achieved with 63\% probability. Inference from our film quality classifier on the deposited film matched this prediction and was verified by visual inspection of the deposited film RHEED. The same RHEED features across two angles were given to the compositional regression model, with the two RHEED angle predictions averaging to a predicted doping level of x=0.09. Subsequent XPS analysis for testing gave a manual composition of x=0.11. Points are plotted in conjunction with the original training and testing data in Fig. 3c. Without modifying any data infrastructure, all of the predictions were made within 15 seconds of RHEED capture, indicating this framework can provide real-time feedback within the timescales of dynamic advanced materials synthesis. The composition predictions are reflected in the red square sample in Figure 3a.

\section{4. XPS Labeled Data for Vanadium Doping}

\begin{figure}[!htb]
    \includegraphics[width=0.5\linewidth]{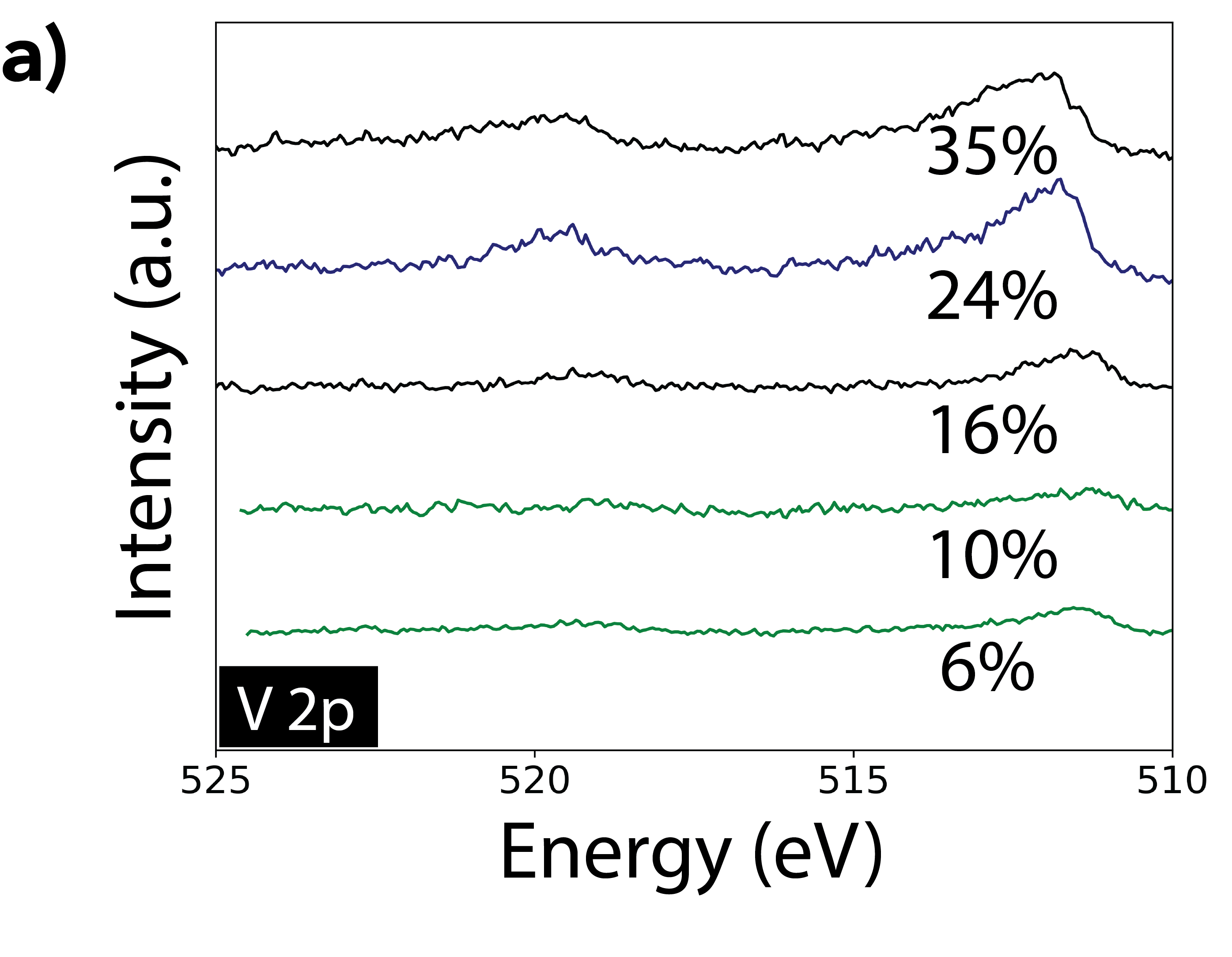}
    \caption{(a) XPS data used to generate the labels for each sample of \ce{W_{1-x}V_xSe2}; percentages are labels extracted using conventional XPS peak-fitting software for the value of $x*100$.}
    \label{figureS3}
\end{figure}

\section{5. RHEED Composition Prediction Model - Metric Correlation Coefficients }

\begin{figure}[!htb]

    \includegraphics[width=0.75\linewidth]{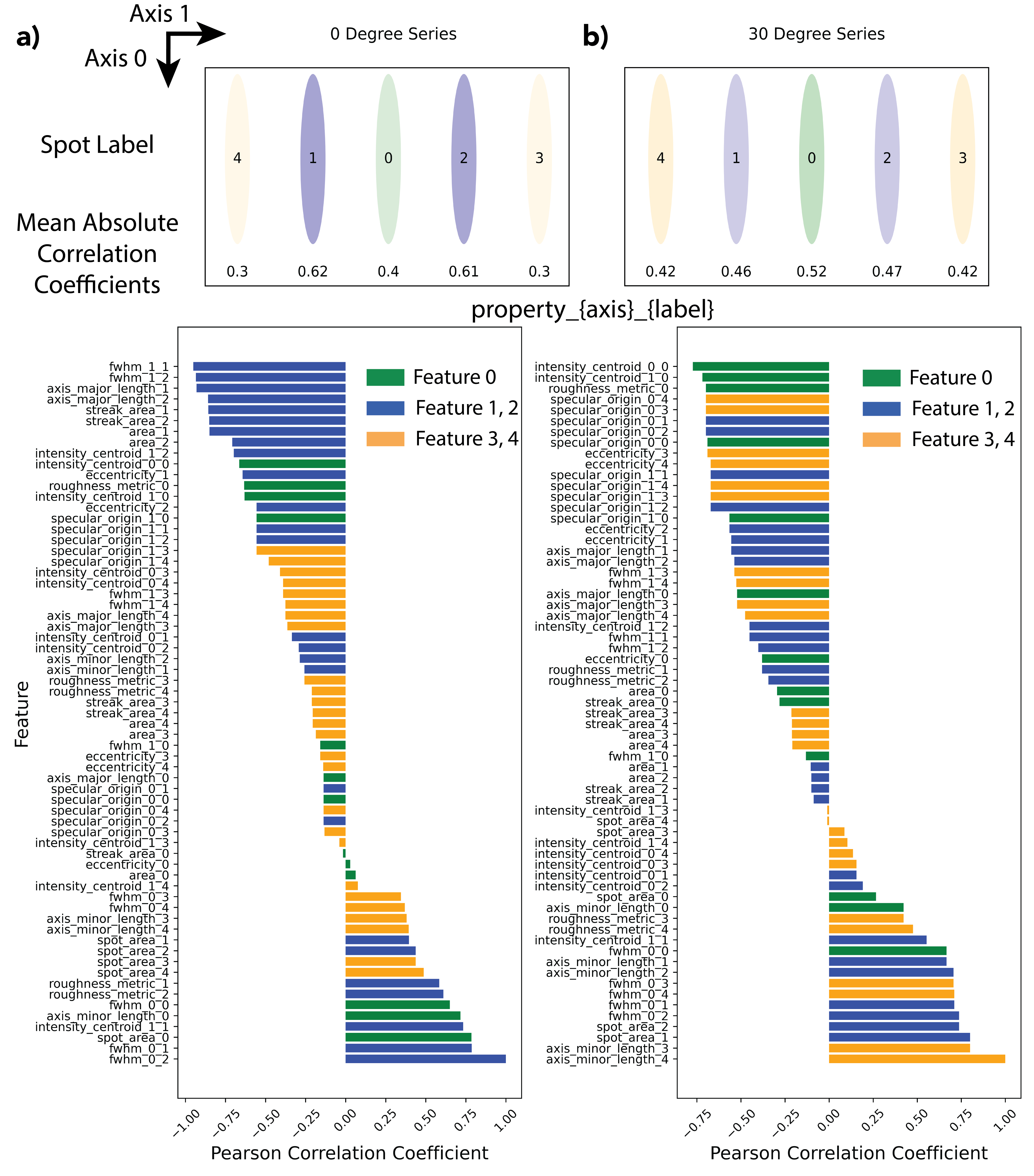}
\caption{(a) Visual feature diagram for the 0-degree azimuthal data series (top) and Pearson correlation coefficients for each metric included in the set of pattern fingerprint attributes. The naming convention is property-axis-label, where axis indicates vertical (0) or horizontal (1) direction and label corresponds to the spot labels. Correlation coefficients of -1 or 1 indicate perfect negative or positive linear correlation between the feature and the target dependent variable (V-doping composition). Bars are color-coded by  label to indicate groupings of specular, primary, and secondary features. The mean absolute correlation coefficients, calculated as the average magnitude of metric correlation grouped by feature, display horizontal symmetry about the specular spot, as is physically expected. (b) Same as (a) for the 30-degree rotated azimuthal data series.}
    \label{figureS4}
\end{figure}

\clearpage
\bibliography{manuscript}